\documentclass[runningheads]{llncs}

\usepackage{graphicx}
\usepackage{textcomp}
\usepackage{xcolor}
\usepackage{hyperref}
\usepackage{booktabs}
\usepackage{lipsum}
\usepackage{comment}
\usepackage{dependencies}
\usepackage{tabularx}
\usepackage{rotating}
\usepackage{svg}

\usepackage{blindtext}

\usepackage{algorithm,algpseudocode}
\usepackage{amsmath}

\begin{document}

\title{Empirical and Theoretical Analysis of \\Liquid Staking Protocols}

%
%

\author{Krzysztof Gogol\inst{1} \and
Benjamin Kraner\inst{2}\and
Malte Schlosser\inst{3}\and
Tao Yan\inst{2}\and \\
Claudio Tessone\inst{2}\and
Burkhard Stiller\inst{1}
}

\authorrunning{K. Gogol, et al.}
%
\institute{Communication System Group, Department of Informatics, University of Zurich\and
Blockchain and Distributed Ledger Technologies Group,\\ Department of Informatics, University of Zurich \and
Department of Banking and Finance, University of Zurich}


%
\maketitle              
\begin{abstract}
 Liquid staking has become the largest category of decentralized finance protocols in terms of total value locked. However, few studies exist on its implementation designs or underlying risks. The liquid staking protocols allow for earning staking rewards without the disadvantage of locking the capital at the validators. Yet, they are seen by some as a threat to the Proof-of-Stake blockchain security. 
 
 This paper is the first work that classifies liquid staking implementations. It analyzes the historical performance of major liquid staking tokens in comparison to the traditional staking for the largest Proof-of-Stake blockchains. Furthermore, the research investigates the impact of centralization, maximum extractable value and the migration of Ethereum from Proof-of-Work to Proof-of-Stake on the tokens' performance. Examining the tracking error of the liquid stacking providers to the staking rewards shows that they are persistent and cannot be explained by macro-variables of the currency, such as the variance or return. 

\end{abstract}

\section{Introduction}

\begin{table*}[ht]
\centering
\caption{Top 10 Crypto Assets by Staking Marketcap\cite{2023StakingRewards}, \cite{2022DeFiLlama}}
\begin{tabularx}{\textwidth}{XXXcccc} 
    \textbf{BC} & \textbf{Token} & \textbf{Lockup} & \textbf{Reward} & \textbf{Adj. Reward} & \textbf{MCap} & \textbf{LSPs} \\
    \toprule 
    Ethereum\cite{Buterin2014Ethereum:Platform.} & ETH & $\infty$ & 4.82\% & 5.04\% & \$35.8bn & 21 \\
    \hline
    Cardano\cite{2023Cardano} & ADA & 0 days & 3.25\% & 0.14\% & \$9.2bn & 0 \\
    \hline
    Solana\cite{2022Solana} & SOL & 1-3 days & 6.51\% & -1.03\% & \$8.1bn & 9 \\
    \hline
    BNB Chain\cite{2022BinanceChain} & BNB & 90 days & 2.69\% & 8.27\% & \$7.6bn & 8 \\
    \hline
    Avalanche\cite{2023Avalanche} & AVAX & 14 days & 7.94\% & 2.23\% & \$4.6bn & 5 \\
    \hline
    Polygon\cite{2023Polygon} & MATIC & 3-4 days & 6.24\% & 3.37\% & \$4.0bn & 4 \\
    \hline
    Polkadot\cite{WoodG.2020Polakdot:FRAMEWORK} & DOT & 28 days & 14.35\% & 6.67\% & \$3.8bn & 1 \\
    \hline
    Tron\cite{2023Tron} & TRX & 3 days & 3.78\% &  1.7\% & \$2.6bn & 2 \\
    \hline
    Cosmos Hub\cite{KwonJ.2020CosmosLedgers} & ATOM & 21 days & 22.78\% & 4.2\% & \$2.5bn & 1 \\
    \hline
    Internet Computer\cite{2023InternetComputer} & ICP & 180 days & 7.4\% & -2.39\% & \$1.8bn & 0 \\
    \bottomrule  
    \end{tabularx}
\label{tab:StakingRwards}
\end{table*}

Ethereum \cite{Buterin2014Ethereum:Platform.} successfully migrated from Proof-of-Work (PoW) to Proof-of-Stake (PoS) consensus mechanism in 2022. After the Shanghai upgrade in 2023, it became possible to unstake Ether (ETH) \cite{2023EthereumRoadmap}. Within the Decentralized Finance (DeFi) space, staking is seen as a relatively safe token allocation that generates steady returns. The inherent disadvantage of the process is the necessity to lock tokens for days or months, depending on the blockchain (BC)\ref{tab:StakingRwards}. \emph{Liquid Staking Protocols} (LSPs) manage to overcome this disadvantage by constructing synthetic (pegged) tokens - \emph{Liquid Staking Tokens} (LSTs) that accumulate rewards from the staking process without the need to lock any assets. Lido \cite{2020Lido:Whitepaper}, the first LSP, became the largest DeFi protocol in terms of total value locked (TVL)\cite{2022DeFiLlama}. Liquid staking has become the largest DeFi category with 78 protocols and more than 16 billion USD TVL\cite{2022DeFiCategories}. LSPs vary in terms of implementation design, level of decentralization, governance, and the validators' selection process. MEV (Maximal Extractable Value)\cite{Chitra2022ImprovingRedistribution}\cite{Qin2021QuantifyingForest} attacks are performed by validators who reorder transactions in the block in order to generate profit. Some LSPs include the earning from MEV attacks in the LSTs rewards.

The benefits of LSPs surpass just unlocking the staked assets. By making the tokens liquid, LSPs allow to fully benefit from DeFi composability ("DeFi Lego") - compounding the returns from various DeFi protocols. LSTs can be additionally allocated to other DeFi protocols. The typical LST allocation strategy involves the liquidity pools in DEXs (Uniswap\cite{Adams2021UniswapCore}, Balancer\cite{MartinelliF2019BalancerSensor.}) or interest rate protocols (Aave\cite{2020AaveV1.0}, Compound\cite{RobertLeshner2019Compound:Protocol}).

LSTs belong to the broader class of DeFi protocols of synthetic (pegged) tokens. DeFi synthetic tokens are digital assets that maintain the peg to the target value. The best-known synthetic tokens are stablecoins. Stablecoins maintain the peg to the fiat currency, typically one US Dollar, \eg MakerDAO\cite{TheMakerTeam2017MakeDAOWhitepaper}, Tether \cite{2023Tether}. Other types of synthetic assets, wrapped tokens, provide interoperability between blockchains. Wrapped Bitcoin\cite{2023WrappedBitcoins} or Wrapped Solana\cite{2022Solana} allows exposure to native tokens of Bitcoin\cite{Nakamoto2008Bitcoin:System} and Solana \cite{2022Solana}, respectively, on the Ethereum blockchain.
The major risk related to synthetic assets is the \emph{de-peg risk} - the risk that a synthetic token loses the peg to the target value. The depegs might be temporary, \eg USDC\cite{2023CircleCoin} or DAI\cite{TheMakerTeam2017MakeDAOWhitepaper}, or permanent, UST\cite{2023Terra}, leading to the failure of the protocols. The collapse of UST, the algorithmic stablecoin in 2022, resulted in the collapse of its underlying blockchain and the entire ecosystem. Liquid staking tokens are exposed not only to the slashing risk, related to staking, but also, as synthetic assets, to the depeg risk.

\paragraph{Methodology and Contribution}
Whereas liquid staking becomes the largest in terms of TVL DeFi category \cite{2022DeFiCategories}, there is very little attention is paid to the matter in academic research. The first and only paper \cite{Scharnowski2022LiquidDiscovery}, examines mainly the stability of the Lido peg on the Ethereum and Solana blockchains. 
This paper makes the following contributions.
\begin{itemize}
\item It provided the taxonomy of the staking reward distribution currently employed by the LSPs, resulting in three categories: rebase, reward, and dual model. 
\item It analyzed data on market values and reserves of major LSTs that represent 90\% of capital in liquid staking, and presented the first comparative study of the historical returns of LSTs against native staking from the three largest PoS blockchains - Ethereum, Solana, and Binance Smart Chain, finding that market values of LSTs are affected by market events, such as Terra/Luna crash or FTX insolvency.
\item It showed, by examining the liquid stacking providers to the staking rewards, that tracking errors are persistent and cannot be explained by macro-variables of the currency, such as the variance or return.
\end{itemize}

\section{Background}

This section introduces the key concepts behind blockchain technology, \eg consensus mechanism, Proof-of-Stake (PoS) consensus mechanism and staking. It further defines synthetic tokens, a class of DeFi protocols to which LSPs belong, and explains the token price mechanisms and the difference between the market and the peg values.

\paragraph{Blockchain and Proof-of-Stake}
A blockchain is a computer network that maintains a distributed database. This network only allows the appending of new transactions and forbids the deleting or updating of any transactions. Transactions are grouped into blocks, and the order of blocks in the database is maintained and verifiable using cryptographic hash functions and digital signatures. For this paper, we define a blockchain as a public, trustless, and permissionless distributed ledger.
A \emph{consensus mechanism} defines the rules for what constitutes a legitimate transaction and a block. The dominant consensus mechanisms are PoW or PoS. The consensus mechanism, based on game theory, economically incentives network participants - miners in PoW and validators in PoS blockchains - to promote network security. In PoW BCs, miners, to append a new block to the ledger, are required to solve the cryptographic puzzle in a process that requires electricity\cite{Nakamoto2008Bitcoin:System}. In PoS BCs, validators, to append a new block, must pledge ("stake") a certain amount of native tokens that act as collateral. In exchange for adding a new block, validators receive a reward. In case of malicious behavior, decreased up-time or latency validators are punished in a process called \emph{slashing} by a fee from the staked tokens. \emph{Staking} is the process of providing native BC tokens to validaters to participate in profits from the block rewards. By design, the PoS blockchain requires validators and \emph{stakers} to freeze their tokens for a period of time that may vary from days to years. \emph{Gas fee} is paid to miners and validators for appending the transaction to the BC. 

\emph{Ethereum Migration to PoS:}
On September 15, 2022 Ethereum completed its transition from PoW to PoS consensus mechanism, also known "the Merge" upgrade. The process started in December 2020, with the launch of staking and the Beacon Chain, and finished in April 2023, with the "Shanghai" upgrade. As of "Shanghai" upgrade, it is possible to unstake the Ether \cite{2023EthereumRoadmap}. 

\emph{PoS Blockchains:}
The comparison between the top 10 PoS BCs in market capitalization is presented in table \ref{tab:StakingRwards}. The table also presents the staking rewards and lockup periods, among others. Until the "Shanghai" upgrade in April 2023, it was not possible to unstake the Ether. After the upgrade, the unstaking process lasts ca 1-2 days, similarly to Solana chain. Currently, there are no liquid staking protocols in Cardano, the second largest BC, as it offers the automatic staking of all tokens held by a BC address without any lock-up period. 


\begin{table}[t!]
\centering
\caption{Synthetic tokens and their target (peg) values}
\begin{tabularx}{\columnwidth}{XX} 
    \textbf{Synthetic Token} & \textbf{Target}\\
    \toprule 
     Stablecoins & Fiat Currency, Commodity\\
    \hline
     Wrapped Tokens & Cryptocurrency\\
    \hline
     Liquid Staking Tokens & Staked Token\\
    \bottomrule  
    \end{tabularx}
\label{tab:SyntheticTokens}
\end{table}

\paragraph{Synthetic (Pegged) Tokens}
Synthetic (pegged) tokens in DeFi have their value pegged to the target values, typically fiat currencies, commodities, or other cryptocurrencies \cite{Rahman2022SystematizationManagement}. 
They track the price fluctuations of the target value\cite{Rahman2022SystematizationManagement}, \cite{BajpaiP2022CryptocurrencyExchanges} and their most well-known examples include stablecoins and wrapped tokens. Stablecoins are pegged to fiat currency, typically 1 USD. Wrapped tokens provide blockchain interoperability by having a peg to tokens minted at other chains; \eg WrappedSolana on the Ethereum blockchain is a synthetic token with a peg to SOL - the native token Solana blockchain. LSTs are synthetic tokens that peg the value to the staked native tokens and their accumulated staking rewards. 

\emph{Fair and Market Value—}
There are various algorithms that synthetic tokens use to maintain the peg to the target value. The most common is holding the reserves of the target assets. This approach ensures that each synthetic token has a potential redemption value, also known as the \emph{peg value}. The fair value is calculated by dividing the value of the reserves by the number of tokens in circulation. 
However, tokens cannot always be freely exchanged at this peg value. For example, stETH from Lido and cbETH from Coinbase have only introduced this possibility since the Shanghai upgrade in April 2023, which makes unstaking in the BC possible.
The \emph{market value} of synthetic tokens is the value at which tokens are traded on decentralized exchanges (DEX) or centralized exchanges (CEX). The difference in value between the fair and market value of synthetic tokens may be caused by market inefficiencies and may result in arbitrage opportunities. 

\section{Liquid Staking Protocols}
\emph{Liquid staking} is a tokenized representation of staked assets in PoS blockchains, and consequently, its value is pegged to the staked tokens \cite{Scharnowski2022LiquidDiscovery}. Liquid staking tokens (LSTs) are also referred to as \emph{Liquid Staking Derivatives} (LSDs), whereas the protocol that mint LSTs are \emph{Liquid Staking Protocols} (LSPs).
Staking is the process of locking tokens in validators to support the PoS blockchain security and, in exchange, earn profits \cite{Scharnowski2022LiquidDiscovery}. Validators, network participants on the PoS blockchain, put those tokens at stake when appending or attesting new transaction blocks to the blockchain. The validator receives a reward for each correctly appended block, which is redistributed between token providers (\emph{stakers}). Dishonest behavior is punished by a reduction in staked tokens (\emph{slashing}). By design, the staked tokens are locked for a specific period, making it impossible to use them in other applications, reducing the token liquidity. 
By tracking the value of staked assets, liquid staking tokens allow participating in the staking benefits. Unlike staked tokens, liquid staking tokens are not locked but are freely traded and can be used in DeFi applications. 

\begin{algorithm}[H]
\caption{Minting LST by LSP}
\label{alg:minting}
\begin{algorithmic}[1]
\State \textbf{Collect} native tokens from the Depositor.
\State \textbf{Stake} deposited native tokens in the LSP's validators.
\State \textbf{Mint} new LSTs, with the value of native tokens deposited, and provide LSTs to the depositor.
\end{algorithmic}
\end{algorithm}

\begin{algorithm}[H]
\caption{Burning LST by LSP}
\label{alg:burning}
\begin{algorithmic}[1]
\State \textbf{Collect} LSTs from the Depositor that redeems LSTs for the native tokens.
\State \textbf{Unstake} native tokens with the value of collected LSTs and provide unstaked native tokens to the Depositor.
\State \textbf{Burn} the collected LSTs.
\end{algorithmic}
\end{algorithm}

\begin{algorithm}[H]
\caption{Arbitrage Strategy when market value $<$ fair value}\label{alg:arbitrageLow}
\begin{algorithmic}[1]
\State \textbf{Buy} undervalued LSTs at DEX for the market value.
\State \textbf{Burn} LSTs at LSP for the peg value.
\end{algorithmic}
\end{algorithm}

\begin{algorithm}[H]
\caption{Arbitrage Strategy when market value $>$ peg value}
\label{alg:arbitrageHigh}
\begin{algorithmic}[1]
\State \textbf{Mint} LSTs at LSP for the peg value.
\State \textbf{Sell} LSTs at the DEX for the market value.
\end{algorithmic}
\end{algorithm}

Blockchain users have two possibilities to acquire LST: \i buy LSTs at DEX or CEX for the market price or \ii mint LSTs directly at the LSP for a fair price. LST minting and burning processes are presented by algorithms \ref{alg:minting} and \ref{alg:burning}, respectively. The blockchain user that mints the LSTs is called \emph{Depositor}, as he deposits the native tokens in the LSP in exchange for the LSTs. The value of the newly minted LSTs is equal to the value of the deposited tokens, so the fair value remains unchanged. LSP distributes the native tokens deposited to the validators for staking. There are three models of distributing the staking rewards to the holders of LSTs: \1 rebase, \2 reward, and \3 dual model - these models are further described in the next section. 
Similarly to acquiring LSTs, the blockchain user has two options to redeem the LSTs: \1 sell LSTs at DEX or CEX for the market price or \2 burn LSTs directly at the LSP for a fair price. When the LSTs are redeemed at the LSPs, the equivalent value of native tokens is unstaked and LSTs are burned.

\emph{Arbitrage Strategies—}
Market inefficiencies could lead to a difference between the fair and market value of synthetic tokens. Algorithms \ref{alg:arbitrageHigh} and \ref{alg:arbitrageLow} present the arbitrage strategies that equal the fair and market values of the LSTs. When the fair value of LSTs is higher than the market price, arbitrageurs buy the LSTs at DEX for the market value and burn at LSP for the fair value. In the opposite circumstance, the arbitrageurs mint the LSTs at LSP for the fair value and sell at DEX for the market value, which is higher than the fair value. 

\emph{Risks—}
The \emph{de-peg risk}, the major risk related to synthetic tokens, including LSTs, is the risk of the token losing the peg to the target value. Depeg might be caused by vulnerabilities in the technical or economical design of the smart contract for synthetic tokens and can be exploited in hacking attacks or oracle price attacks.
In addition to the de-peg risk, LSTs are vulnerable to the slashing risks caused by the LSP validators. 

\section{Taxonomy of Liquid Staking Protocols}

This section presents the taxonomy of LSP. The taxonomy is straightforward, as all information can be directly retrieved from the whitepaper or the source code of the LSPs.

\begin{figure*}[h]
\centerline{\includegraphics[width=1.2\textwidth]{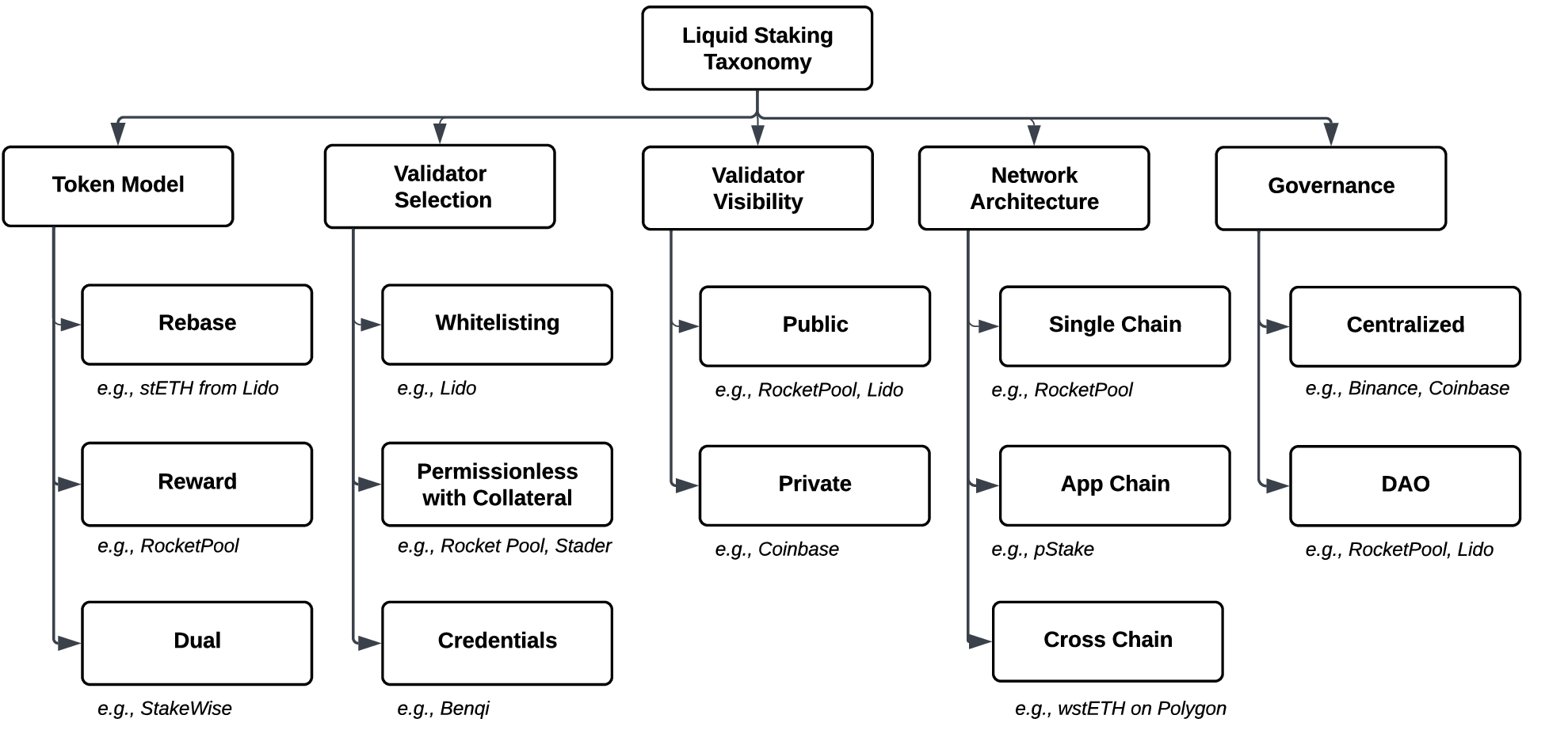}}
\caption{Taxonomy of Liquid Staking Protocol}
\label{fig:taxonomyLSP}
\end{figure*}

\subsection{Taxonomy by Token Model}
The major differentiation between LSPs is the mechanisms for distributing the staking rewards. LSPs apply three standard models for this purpose \cite{2022AProtocols:}: \i rebasing tokens, \ii reward-bearing tokens, and \iii3 dual token model.

\emph{Rebase Tokens—} In this model, LSP tokens keep the 1:1 connection to staked tokens and inflate the supply to reflect the profit from stake. The most popular rebase LSP is stETH, minted by Lido Protocol\cite{2020Lido:Whitepaper}. Lido every day increases the balance of all stETH holders to reflect the additional ETH earned as staking rewards. The advantage of rebasing LSP is its simple functionality. The disadvantage is the lack of compatibility with other DeFi protocols, \eg DEXs, or interest rate protocols. 

\emph{Reward-bearing Token—} In this model, the token increases its value to reflect the staking yields and does not maintain a 1:1 peg to the native token. Consequently, the reward-bearing token grows in value related to the underlying native token. Reward-bearing tokens provide full compatibility with DeFi protocols, and consequently, most LSPs follow this design: RocketPool\cite{2023RocketPool}, Marinade \cite{2023Marinade}, CoinBase Wrapped Staking \cite{2022CoinBaseWhitepaper}.

\emph{Dual Token—} In the dual token model, the asset token is split from the stake revenue it generates. For example, StakeWise \cite{2023StakeWise} operates in the two-token model: sETH2 - with a 1:1 peg to ETH -  and rETH2 - which reflects the earned staking rewards. The disadvantage of this model is the fragmented liquidity between two tokens.  

\subsection{Taxonomy by Validator Selection} LSP also varies in the validator selection process: \i whitelisting reputable validators, \ii credential-based, \iii collateral. In the whitelisting model, LSP adds only trusted validators to the pool of validators, \eg Lido\cite{2020Lido:Whitepaper}. In the credential-based model, the pool of validators is permissionless after fulfilling the minimum criteria. In the last model, LSP requires validators to post collateral to guarantee its performance, \eg Rocket Pool\cite{2023RocketPool}. The collateral model hampers the growth of the validator pools by increasing the minimum threshold for staking. 

\subsection{Taxonomy by Network Architecture}
LSPs can operate in a single chain or in multiple chains. Single-chain protocols are deployed and operate within one BC, \eg. Rocket Pool\cite{2023RocketPool} on Ethereum. App-chain (or Solo-chain) is LSP that developed sovereign BC to optimize its services, \eg pStake\cite{2023PSTAKE}.
Cross-chain protocols operate on multiple BCs and allow the value to cross between BCs, \eg Lido allows one to purchase stETH on the Polygon BC. 

\subsection{Revenue Models and Token Utility}
The most common revenue model for LSP is commission on staking yields with the standard fee is 10\%. The token utility includes \i governance \ii revenue sharing \iii validator collateral.
\emph{Governance Utylity-} Holder of the LSP utility tokens participates in the DAO votes.
\emph{Revenue Sharing-} LSPs charge a commission for their services and collect it in the protocol treasury. LSP utility token holders can participate in the distribution of part of the revenues from the treasury.
\emph{Validator Collateral-} LSPs might require to deposit collateral in their utility tokens.
RocketPool\cite{2023RocketPool} enabled permissionless validation selection by requiring validators to post their utility tokens - RPL tokens - as collateral. The deposited collateral is also used as protection against slashing.

\section{Data}
The analysis is performed within the three largest PoS blockchains regarding market cap: Ethereum\cite{Buterin2014Ethereum:Platform.}, Solana\cite{2022Solana}, and BNB Chain\cite{2022BinanceChain}.
First, we analyze the performance of the LSTs in comparison with the staking of the corresponding native tokens ETH, SOL and BNB. We calculated the daily performance of LSTs based on the market price we sourced directly from UniSwap\cite{Adams2021UniswapCore} for LSPs on Ethereum and from Yahoo Finance for others. The reason we do not use the peg price is that we model the realistic token allocation strategy that might be performed by the DeFi user.For each analysis, we consider one unit of native token that can be allocated to LSTs or into staking. When calculating staking rewards, we assume for simplicity that the rewards are claimed daily and re-staked. Staking yield rates change every day and vary between validators. Staking rates are non-linear and are decided in the blockchain governance process by Ethereum, Solana and Binance. We use the exact historical staking rates from the Ethereum explorers\cite{} and assume flat rates for Solana and Binance. \\

\begin{figure*}[!tbp]
  \centering
  \begin{minipage}[b]{\textwidth}
    \includegraphics[width=\textwidth]{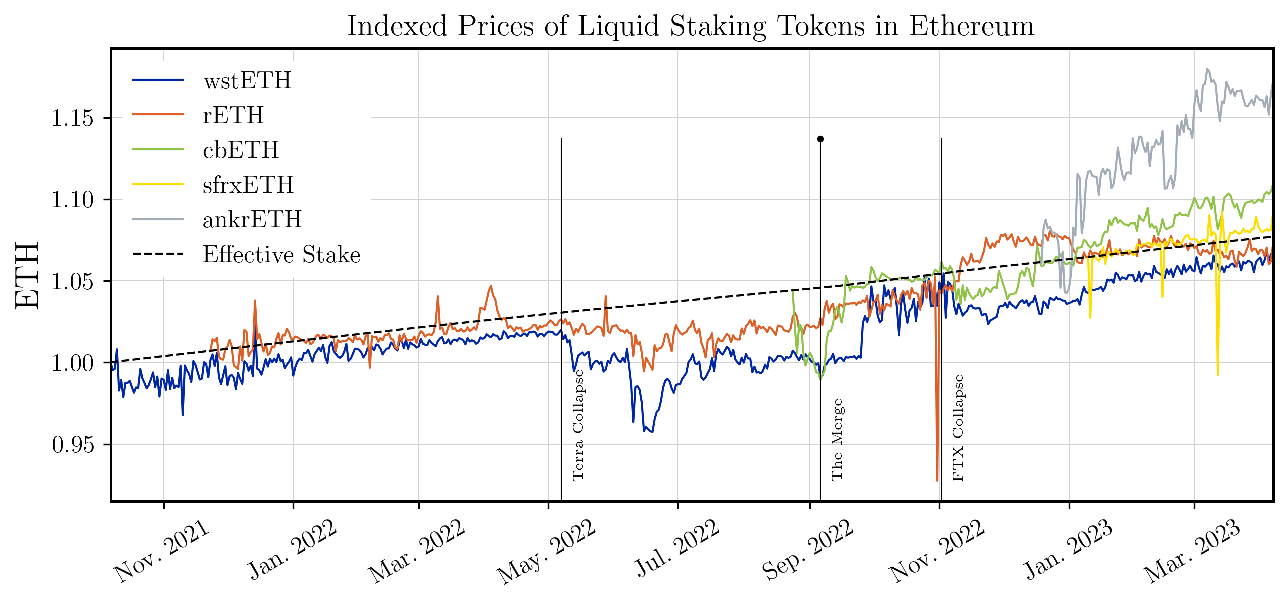}
    \caption{Comparison of selected LSTs with staking ETH [ETH]}
    \label{fig:Ethereum}
  \end{minipage}
  \hfill
    \begin{minipage}[b]{\textwidth}
    \includegraphics[width=\textwidth]{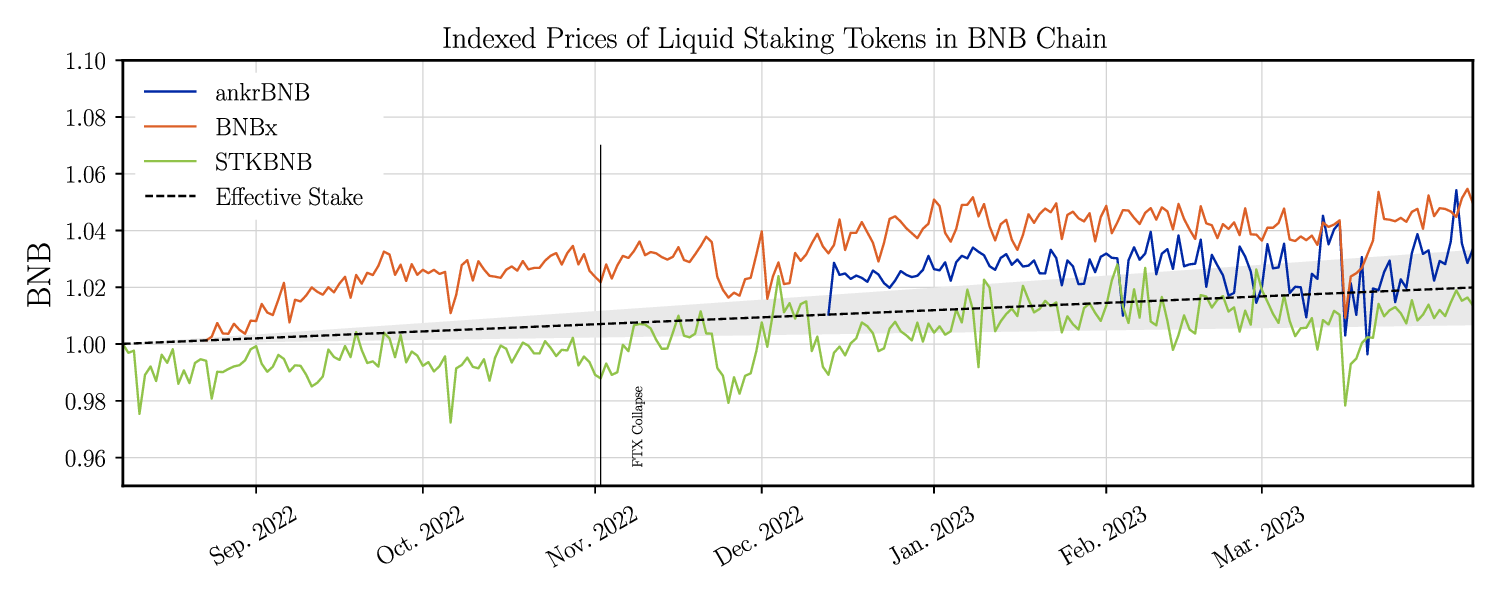}
    \caption{Comparison of selected LSTs with staking BNB [BNB]}
    \label{fig:Binance}
  \end{minipage}
  \hfill
  \begin{minipage}[b]{\textwidth}
    \includegraphics[width=\textwidth]{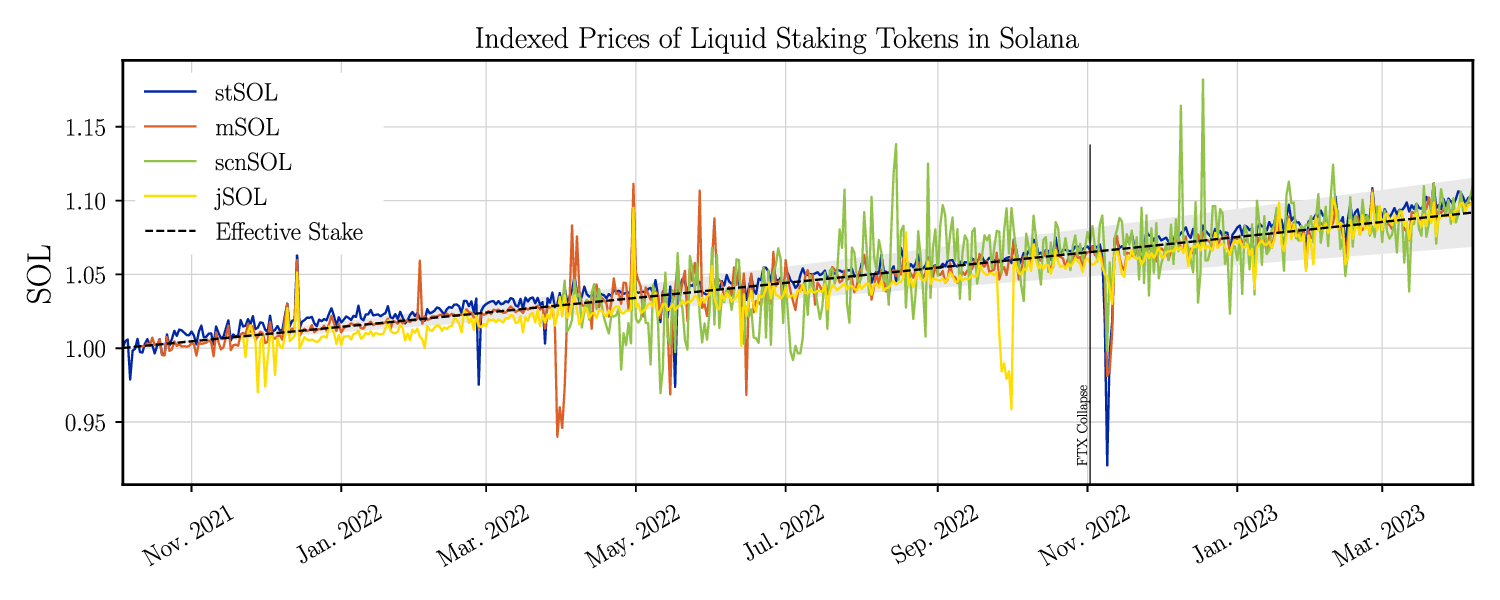}
    \caption{Comparison of selected LSTs with staking SOL [SOL]}
    \label{fig:Solana}
  \end{minipage}
\end{figure*}

\begin{figure}[!tbp]
  \centering
    \includegraphics[width=\textwidth]{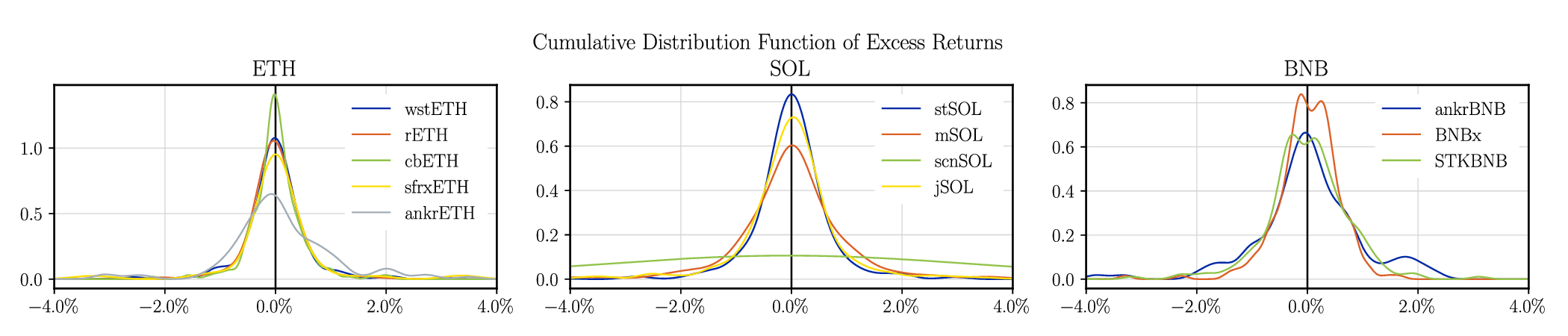}
    \caption{Cumulative distribution function of excess returns (in percent) of LSP tokens in Ethereum, Solana and BNB Chain. }
    \label{fig:d_Ethereum}
\end{figure}

\subsection{Ethereum}
The analysis includes reward-bearing tokens in the Ethereum chains. This approach allows a comparison of how effectively the market price of LSTs accumulates the ETH staking rewards. The value of re-base LSTs, such as stETH or frxETH, tracks the value of 1 ETH and the staking rewards are distributed daily to the token holder as additional tokens. Thus, tracking their value does not allow to compare the accumulated rewards from staking. LSPs, such as Lido or Frax, issue two LSTs: re-based and reward tokens, stETH and wstETH and frxETH and sFRX, respectively. The reward tokens analyzed include wstETH Lido, rETH Rocket Pool\cite{2023RocketPool}, cbETH Coinbase\cite{2022CoinBaseWhitepaper} and sfrxETH Frax\cite{2023Frax}. Figure \ref{fig:Ethereum} presents the compounded daily returns of the reward-bearing LSTs on the Ethereum blockchain against staking Ethereum. 

\emph{Dispersion Analysis—}
The dispersion analysis of ETH LSTs is presented in figure \ref{fig:d_Ethereum}. The figure presents the distribution of the excess return, calculated as the absolute difference between the daily LST returns and the daily returns from ETH staking. The fees charged by LSPs lead to a negative shift in the distribution, the inclusion of MEV rewards in a rightward shift. Eventually, the LSP returns are slightly negative, as indicated by a negative median return. On a daily return basis, cbETH has the lowest tracking error, while the highest deviation occurs for ankrETH tokens. Table \ref{table5_disp} confirms this, since the standard deviation of the excess return is the lowest for cbETH and all tokens have a negative median. 

\emph{Terra and FTX collapses—}
The crypto market was severely impacted in 2022 by adverse market situations. The first was the crash of the algorithmic USDT stablecoin that caused the collapse of Terra BC and its entire ecosystem of protocols (7 May 2020). Following the collapse of Terra, the centralized exchange (CEX) - FTX - became insolvent (2 November 2022), freezing the digital assets of its customers. In the period between the collapse of Terra and FTX, both the market values of the stETH and rETH tokens performed worse than traditional staking. In the aftermath of the FTX collapse, the rETH token performed better than stake, which can be attributed to the investor's reluctance to centralized protocols such as FTX. Rocket Pools, due to its permissionless selection of validators, is considered more decentralized than the Lido protocol, which whitelists its validators. However, to protect against dishonest validators, Rocket Pool requires validators to deposit the collateral. The drop in the market value of LSTs might be caused by \i market inefficiencies in pricing the token or \ii drop in the value of its reserves (\eg as a result of slashing or hacking). When the LST's market value differs from the peg value, the market is inefficient, and arbitrage exists. 

\begin{figure*}[h]
\centerline{\includegraphics[width=\textwidth]{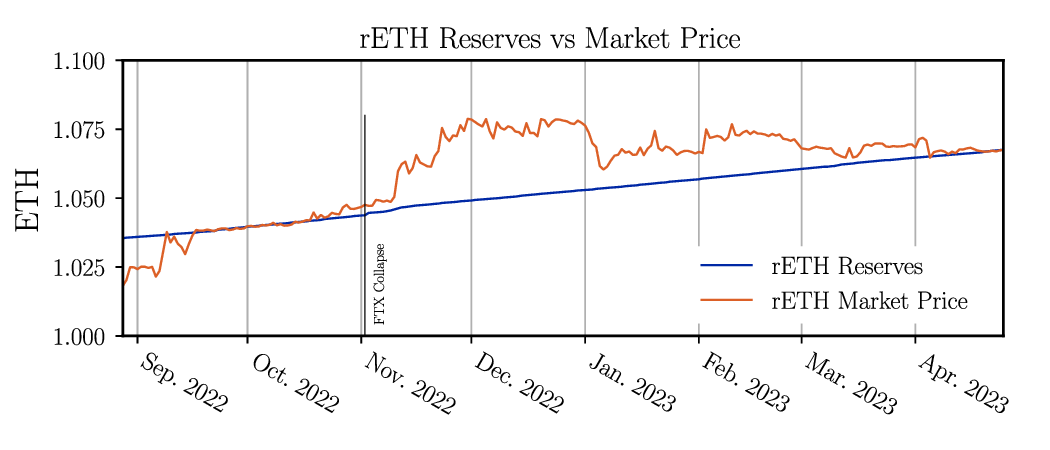}}
\caption{Pegged value vs market value for rETH}
\label{fig:peg_rETH}
\end{figure*}

\emph{Peg Analysis—}
Until the Shanghei upgrade, there was no possibility to unstake ETH. Consequently, the market value of LSTs pegged to staked ETH might deviate from their peg value. Figure \ref{fig:peg_rETH} illustrates the market and peg values for the rETH token of RocketPool. In the period between the Terra-Luna collapse (7 May 2022) and FTX collapse (Nov 2, 2022), the rETH was underpriced compared to its peg value. If unstaking ETH would be possible in this period, arbitrageurs could equal the market and peg values of rETH. Following the FTX crash and until January 2023, the rETH tokens were overpriced, and the arbitrage strategy existed, algorithm \ref{alg:arbitrageHigh}. The rETH could be minted at the Rocket Pool Protocols for the peg value and sold at the DEX (UniSwap) for the market value, which was higher than the peg value. The over-pricing of LSTs from Rocket Pools can be attributed to the reluctance to the centralized protocol following the FTX collapse. 

\subsection{Solana}
We analyze the following LSPs in Solana BC - stSOL from Lido\cite{2020Lido:Whitepaper}, mSOL from Marinade\cite{2023Marinade}, jSOL from JPool\cite{2023JPool} and scnSOL from Socean\cite{2023Socean.fi}. All tokens are reward-bearing and their value accumulates daily staking rewards from staking SOL - the native token of the Solana BC. All LSPs are deployed to the Solana BC. Figure \ref{fig:Solana} presents the compounded daily returns of these tokens against SOL stakes. All tokens perform similarly, with the temporary drop in value in November 2022 related to the collapse of the FTX exchange (2. Nov) that had a significant involvement in the Solana BC and its ecosystem. Following the Terra collapse, market values of LSTs on Solana were not traded undervalued like ETH-based tokens, which can be explained by the possibility to unstake SOL within days, compared to the lockup of ETH in staking until the Shangai upgrade. The dispersion analysis - figure \ref{fig:d_Ethereum} - shows that all LSTs on Solana track the SOL staking rewards.

\subsection{BNB Chain}
We analyze the performance of LSTs tracking the staking rewards of native token BNB on the Binance Chain: Ankr\cite{2023Ankr}, Stader\cite{2023StaderLabs}, pStake\cite{2023PSTAKE}. All tokens are reward-bearing and accumulate the staking rewards daily in the token value. Figures \ref{fig:Binance} and
\ref{fig:d_Ethereum} present compounded daily returns of these tokens against BNB staking, and their dispersion from the BNB staking returns, respectively. BNB Chains offers 30, 60, and 90 day staking periods, with higher staking rewards for the longer period. With the LSTs on BNB it is possible to benefit from the higher rates.

\section{Empirical Analysis}

\subsection{Model}
We use multivariate regressions with trend coefficients to analyze potential reasons for the deviations of the daily return and the premium/discount. We follow \cite{Hayashi} as a basis for the econometric models.\\

A multivariate regression tests the relationship of several independent variables to a dependent variable with an ordinary least square estimation method. The assumption is that there is a linear relationship between the dependent variable $y$ and the n independent variables $x_1,\dots,x_n$:
$$ y = \beta_0 + \beta_1\cdot x_1 + \beta_2\cdot x_2 + \dots + \beta_n\cdot x_n + \varepsilon,$$ where $\beta_1,\dots,\beta_n$ are the estimation coefficients for $x_1, \dots, x_n$ and $\varepsilon$ is the error term. The parameters $\beta_1,\dots,\beta_n$ are estimated in such a way that the sum of the Euclidean norm is minimised and thus the best possible linear fit is found. \\ 

We use this model to test whether different variables have an impact on the excess return of the LST $(Xs_{LST})$, defined as the difference between the return of the LST and the staked base currency and how big this impact is. We are particularly interested in whether the daily price change, the market capitalization, the volume, or the standard deviation can explain part of the excess returns. The price change is used to test whether stronger revaluations or devaluations of the base currency have an impact on the excess return, which can be caused by positive or negative news about that currency. The market capitalization is used to check if the size of the LSP has an influence on the excess return, while the traded volume is used to analyze if the excess return depends on active or passive days. The standard deviation is used to understand the behavior of the excess return in turbulent market phases. This results in the following model \begin{equation*}
     Xs_c = \alpha + \beta_1\cdot \Delta_{\text{daily}}^c + \beta_2\cdot \sigma_{\text{monthly}}^c + \beta_3\cdot \sigma_{\text{daily change}}^c + \beta_4 \cdot M  \beta_5\cdot  V,
 \end{equation*} where $c$ is a LST on either Ethereum, Solana or BNB, $\Delta_{\text{daily}}$ the daily price percentage price change in USD, $\sigma_{\text{monthly}}$ the standard deviation of the last 30 trading days, $\sigma_{\text{daily change}}$ the daily percentage change in $\sigma_{\text{monthly}}$, MCap represents the market capitalization, $V$ the daily volume, and $Xs_c$ represent the excess returns.\\
  In addition, we use heteroskedasticity-consistent standard errors (HC3) and check the variance inflation factor (VIF) to control for multicollinearity. 

Besides the daily excess return, we want to understand why a premium/discount occurs and whether the above variables can explain when it exists. We have seen in Figures \ref{fig:Ethereum} to \ref{fig:Solana} that the premium/discount has persisted over a longer period of time. Due to this autocorrelation, we cannot perform a simple multiple linear regression and we have to combine it with an autoregressive model to analyze the premium. To determine the lags of the trend component, we analyze the partial auto-correlation function of the premium/discount time series to find out how many previous days have an impact on today's premium. A p-th order autoregressive process AR(p) is defined as  
\begin{equation*}
y_t = \alpha + \beta_1\cdot y_{t-1} + \dots + \beta_p\cdot y_{t-p} + \varepsilon_t, 
\end{equation*} where $y_{t-p}$ is the p-lagged value of $y$ and $\varepsilon_t$ is white noise.
 We find out that for all protocols the effect disappears after 6 days at the latest. Thus, we include a time lag of up to 6 days, which leads to the model:
 \begin{align*}
     \text{Premium}_t = &\alpha + \beta_1\cdot \Delta_{\text{daily}}^c + \beta_2\cdot \sigma_{\text{monthly}}^c + \beta_3\cdot \sigma_{\text{daily change}}^c + \beta_4 \cdot \text{MCap} + \beta_5\cdot V \\  + &\beta_6 \cdot \text{Premium}_{t-1} + \dots + \beta_11 \cdot \text{Premium}_{t-6},
 \end{align*} where Premium is the difference between the market price and staking rate of the LSP and we add 6 lagged trend variables to account for the autocorrelation in the data. By including the lagged variables, we are not only able to see how much the premium/discount depends on the past days, but also to isolate the effect of the other variables from the autocorrelation. \\

\subsection{Results}
Tables \ref{OLSETH} to \ref{OLSBNB} in the appendix show the OLS regressions, with the left half showing the regressions on excess returns and the right half on premium/discount.\\
For daily returns, the BNB and ETH regressions show similiar results. With the exception of BNBx's market cap and volume, none of the factors are relevant at a 5\% level of significance and the alpha cannot be distinguished from 0. This means that most LSPs in ETH and BNB track the staking rewards very well on a daily return level and the smaller deviations are not caused by any of our macro-variables. For BNBx we have a positive correlation between market cap and excess return. This means that the excess return was greater when the market cap of BNBx increased. However, if this has led to a larger turnover of the token, this has offset some of the gains. 
The picture is similar for Solana, except that the $\Delta_{daily}^{SOL}$ factor is significant for all tokens except stSOl. This means that the price development of Solana has been reflected with an increased effect in the price of LSTs. Thus, on days with a positive performance of the Solana-USD price, higher excess returns were earned with the LSTs, while negative price movements of Solana reduced the excess return or led to losses compared to staking. However, even for Solana, the alpha is indistinguishable from 0, which means that the LSTs can track the staking rate very well on a daily basis. \\

Regression of the premium/discount of LSPs yields a different result. The premium strongly depends on the previous values. We see that especially the two previous days show significant positive values.  Thus, the premium tends to persist if there was already a premium on the previous day. Thus, a 1 percentage point higher premium of rETH had a positive impact of 0.491 percentage points on the following day, 0.266 on the second day after and still 0.09 on the third day at a significance level of 1\%. In addition wstETH, sfrxETH and mSOL tend to pay a premium when the underlying currency has a higher market cap and a negative correlation with the volume traded. Apart from scnSOl, an alpha significantly different from 0 was found for all LSTs at a significance level of 1\%. These results imply that the premium persists and is not correlated with the macro-variables of the currency. This is an indication that the discrepancy could be traded away via arbitrage.

\section{Discussion}
LSPs became the largest DeFi category in 2023 in terms of TVL \cite{2022DeFiCategories} and Lido, the first LSP, became the largest in terms of TVL protocol. The new LSPs, such as RocketPool aim to address the centralized nature of Lido. Lido whitelists the selected validators in a DAO vote and centrally manages the rewards, inflows, and withdrawals.  The new LSPs, such as RocketPool, request that the validators deposit collateral to guarantee the honest behaviors of the validators and cover the potential slashing losses. RocketPool, followed by other LSPs, includes in LST rewards from the MEV attacks performed by the validators, but this research found no impact of MEV rewards on the LST's performance. MEV (Maximum Extractable Value) refers to the profit earned by validators by re-ordering the transaction in the block.

\begin{algorithm}
    \caption{Compounding strategy for LST}
    \label{alg:compoundingStrategy}
    \begin{algorithmic}[1]
    \State \textbf{Mint} new LSTs at LSP with native tokens as collateral.
    \State \textbf{Deposit} newly minted LSTs at AMM DEX in the liquidity pool comprising of LST.
    \State \textbf{Deposit} LP-tokens from AMM DEX in the DEX staking mechanism.
    \end{algorithmic}
\end{algorithm}

One of the DeFi promises is its composability, which allows compounding returns from various DeFi protocols. The example DeFi strategy that involves LST, presented in algorithm \ref{alg:compoundingStrategy}, involves allocating LSTs to DEXs. After minting (or purchasing) LSTs, the tokens are locked in the liquidity pool at AMM DEX. The liquidy pool includes the tokens pair: LST and stablecoin. In the last step,  the LP-tokens are allocated to the DEX's staking mechanisms. Other LSTs strategies might involve interest rate protocols (Aave, Compound) to leverage or hedge the LSTs position. DeFi composability allows for a magnifying of the returns of liquid staking.

This work defines staking as an integral part of PoS BC. However, some DeFi protocols offer "to stake" of their governance tokens or LP-tokens (\eg Curve). These tokens are locked for a period of time to accumulate returns from the DeFi protocol fees. 




\section{Summary}
Liquid Staking Tokens (LSTs) are synthetic (pegged) tokens that are tokenized representation of staked assets and consequently track the value of staked tokens and staking rewards. This paper introduces the LST taxonomy, differentiating between the \1 rebase, \2 reward, and \3 dual token implementations. Subsequently, the reward-bearing tokens were empirically analyzed for the largest blockchains in market capitalization: Ethereum, Binance Smart Chain, and Solana.

This research found that LSTs on Solana and BNB Chain blockchains delivered comparable returns to the staking of the native tokens, SOL and BNB, respectively. Ethereum, until the Shanghai upgrade, did not support unstaking ETH, which led to fluctuations in the market prices of LSTs, especially following the Terra collapse and FTX insolvency. 

After the FTX insolvency, the rETH tokens from RocketPool protocol traded overpriced, compared to their peg value, which can be attributed to a reluctance to the centralized protocols. RocketPool is considered to be more decentralized than Lido protocol.


\begin{table*}[t]
\centering
\caption{Overview of analyzed LSPs}
\begin{tabularx}{\textwidth}{XXXXXXXXX} 
    \textbf{LSP} & \textbf{LST} & \textbf{Target} & \textbf{BC} & \textbf{Token Model}\\
    \toprule 
     Lido\cite{2020Lido:Whitepaper} & stETH & staked ETH & Ethereum & rebase\\
    \hline
     Lido\cite{2020Lido:Whitepaper} & wstETH & staked ETH & Ethereum & reward\\
    \hline
     Rocket Pool\cite{2023RocketPool} & rETH & staked ETH & Ethereum & reward \\
    \hline
     \small{CoinBase WSP}\cite{2022CoinBaseWhitepaper} & cbETH & staked ETH & Ethereum & reward\\
    \hline
    Frax\cite{2023Frax} & sfrxETH & staked ETH & Ethereum & reward\\
    \hline
    GETH\cite{2023GETH} & gETH & staked ETH & Ethereum & reward\\
    \hline
     StakeWise\cite{2023StakeWise} & sETH2, rETH2 & staked ETH & Ethereum & dual\\
    \hline
    
     Lido\cite{2020Lido:Whitepaper} & stSOL & staked SOL & Solana & reward\\
    \hline
     Marinade\cite{2023Marinade} & mSOL & staked SOL & Solana & reward\\
    \hline
     Socean\cite{2023Socean.fi} & scnSOL & staked SOL & Solana & reward\\
    \hline
     JPool\cite{2023JPool} & jSOL & staked SOL & Solana & reward\\
    \hline

     Ankr\cite{2023Ankr} & ankrBNB & staked BNB & BNB Chain & reward\\
    \hline
     Stader\cite{2023StaderLabs} & BNBx & staked BNB & BNB Chain & reward\\
    \hline
     pStake\cite{2023PSTAKE} & stkBNB & staked BNB & BNB Chain & reward\\
    \hline
     StaFi\cite{2023StaFi} & rBNB & staked BNB & BNB Chain & reward\\
    
    \bottomrule  
    \end{tabularx}
\label{tab:LiquidStakingTokens}
\end{table*}

\section*{Acknowledgements} 
The Authors thank DIVA and other liquid staking protocols for helpful conversations.

\bibliographystyle{splncs04}
\bibliography{main}

\appendix

\appendix
\setcounter{section}{0}

%
%

\begin{sidewaystable} 
\centering
\small{
\begin{tabular}{lrrrrrrrrrrrr}
\toprule
{} &    wstETH &      rETH &     cbETH &  sfrxETH &   ankrETH &     stSOL &      mSOL &    scnSOL &      jSOL &   ankrBNB &      BNBx &    STKBNB \\
\midrule
Count & 547 & 500 & 226 & 87 & 110& 550 & 542 & 538 & 501 & 116 & 228 & 243 \\
Mean  &   0.00000 &   0.00003 &   0.00014 &  0.00026 &   0.00082 &   0.00007 &   0.00011 &   0.00483 &   0.00007 &   0.00016 &   0.00014 &   0.00000 \\
Std.   &   0.00590 &   0.00920 &   0.00481 &  0.01479 &   0.01000 &   0.01107 &   0.01382 &   0.14620 &   0.01103 &   0.00950 &   0.00561 &   0.00732 \\
Min.   &  -0.03046 &  -0.11865 &  -0.02215 & -0.08002 &  -0.03112 &  -0.09609 &  -0.07849 &  -0.71031 &  -0.06145 &  -0.03821 &  -0.03296 &  -0.03169 \\
25\%   &  -0.00245 &  -0.00214 &  -0.00172 & -0.00127 &  -0.00379 &  -0.00277 &  -0.00462 &  -0.01316 &  -0.00350 &  -0.00374 &  -0.00245 &  -0.00367 \\
50\%   &  -0.00009 &  -0.00013 &   0.00006 & -0.00002 &  -0.00009 &   0.00011 &   0.00011 &  -0.00016 &   0.00011 &   0.00009 &   0.00040 &  -0.00002 \\
75\%   &   0.00239 &   0.00213 &   0.00207 &  0.00159 &   0.00506 &   0.00275 &   0.00429 &   0.01196 &   0.00356 &   0.00480 &   0.00342 &   0.00413 \\
Max.   &   0.03162 &   0.12566 &   0.02045 &  0.07999 &   0.03485 &   0.08239 &   0.07127 &   3.23583 &   0.10835 &   0.02320 &   0.01650 &   0.03097 \\
\bottomrule
\end{tabular}}
\caption{Descriptive Statistics of Excess Returns in different Liquified Staking Tokens}
\label{table5_disp}
\end{sidewaystable}

\begin{sidewaystable} 
\centering
\begin{tabular}{@{\extracolsep{5pt}}lcccccccccc}
\\[-1.8ex]\hline
\hline \\[-1.8ex]
\\[-1.8ex] & wstETH & rETH & cbETH & sfrxETH & akrETH & wstETH & rETH & cbETH & sfrxETH & ankrETH \\
\hline \\[-1.8ex]
 const & 0.000$^{}$ & 0.000$^{}$ & 0.000$^{}$ & 0.000$^{}$ & 0.001$^{}$ & -0.022$^{***}$ & -0.006$^{***}$ & 0.003$^{***}$ & 0.002$^{***}$ & 0.061$^{***}$ \\
  & (0.000) & (0.000) & (0.000) & (0.001) & (0.001) & (0.000) & (0.000) & (0.000) & (0.001) & (0.001) \\
$\Delta^{ETH}_{daily}$ & -0.001$^{}$ & -0.011$^{*}$ & 0.014$^{}$ & 0.032$^{}$ & -0.010$^{}$ & 0.002$^{}$ & -0.008$^{}$ & 0.014$^{}$ & -0.086$^{}$ & -0.012$^{}$ \\
  & (0.005) & (0.006) & (0.012) & (0.034) & (0.038) & (0.005) & (0.006) & (0.013) & (0.077) & (0.040) \\
  $\sigma_{\text{daily change}}^{ETH}$ & -0.005$^{}$ & -0.005$^{}$ & -0.009$^{*}$ & -0.046$^{}$ & 0.005$^{}$ & 0.001$^{}$ & 0.002$^{}$ & -0.007$^{}$ & -0.092$^{}$ & 0.013$^{}$ \\
  & (0.006) & (0.005) & (0.005) & (0.037) & (0.025) & (0.005) & (0.007) & (0.004) & (0.072) & (0.025) \\
 market cap & 0.000$^{}$ & -0.000$^{}$ & 0.000$^{}$ & 0.031$^{}$ & 0.001$^{}$ & 0.002$^{***}$ & 0.002$^{}$ & 0.005$^{}$ & 0.058$^{***}$ & 0.026$^{}$ \\
  & (0.001) & (0.001) & (0.002) & (0.019) & (0.010) & (0.001) & (0.002) & (0.003) & (0.017) & (0.016) \\
  $\sigma_{monthly}^{ETH}$ & 0.003$^{}$ & 0.011$^{}$ & -0.011$^{}$ & -0.277$^{}$ & -0.166$^{}$ & -0.044$^{*}$ & 0.030$^{}$ & -0.030$^{}$ & -0.172$^{}$ & -0.270$^{}$ \\
  & (0.014) & (0.020) & (0.022) & (0.243) & (0.165) & (0.023) & (0.035) & (0.019) & (0.255) & (0.346) \\
 shift1 & & & & & & 0.491$^{***}$ & 0.257$^{**}$ & 0.911$^{***}$ & -0.191$^{*}$ & 0.729$^{***}$ \\
  & & & & & & (0.096) & (0.122) & (0.114) & (0.110) & (0.058) \\
 shift2 & & & & & & 0.233$^{***}$ & 0.266$^{***}$ & -0.056$^{}$ & -0.209$^{***}$ & 0.065$^{}$ \\
  & & & & & & (0.065) & (0.053) & (0.172) & (0.045) & (0.141) \\
 shift3 & & & & & & 0.020$^{}$ & 0.090$^{***}$ & -0.049$^{}$ & -0.012$^{}$ & 0.086$^{}$ \\
  & & & & & & (0.055) & (0.032) & (0.128) & (0.045) & (0.086) \\
 shift4 & & & & & & 0.179$^{***}$ & 0.052$^{*}$ & 0.261$^{***}$ & -0.054$^{}$ & 0.005$^{}$ \\
  & & & & & & (0.051) & (0.031) & (0.085) & (0.079) & (0.111) \\
 shift5 & & & & & & 0.017$^{}$ & 0.166$^{***}$ & -0.134$^{**}$ & 0.001$^{}$ & -0.012$^{}$ \\
  & & & & & & (0.076) & (0.060) & (0.067) & (0.039) & (0.096) \\
 shift6 & & & & & & -0.046$^{}$ & 0.047$^{}$ & 0.004$^{}$ & 0.017$^{}$ & -0.000$^{}$ \\
  & & & & & & (0.052) & (0.047) & (0.056) & (0.037) & (0.073) \\
 volume & -0.000$^{}$ & -0.000$^{}$ & 0.000$^{}$ & 0.010$^{}$ & -0.001$^{}$ & -0.002$^{**}$ & -0.002$^{}$ & -0.000$^{}$ & -0.004$^{}$ & -0.000$^{}$ \\
  & (0.001) & (0.001) & (0.001) & (0.006) & (0.002) & (0.001) & (0.002) & (0.001) & (0.003) & (0.004) \\
\hline \\[-1.8ex]
 Observations & 547 & 500 & 226 & 87 & 110 & 542 & 495 & 221 & 82 & 105 \\
 $R^2$ & 0.005 & 0.003 & 0.026 & 0.058 & 0.007 & 0.876 & 0.682 & 0.946 & 0.320 & 0.873 \\
 Adjusted $R^2$ & -0.005 & -0.007 & 0.004 & -0.001 & -0.041 & 0.873 & 0.674 & 0.943 & 0.213 & 0.857 \\

\hline
\hline \\[-1.8ex]
\textit{Note:} & \multicolumn{10}{r}{$^{*}$p$<$0.1; $^{**}$p$<$0.05; $^{***}$p$<$0.01} \\

\end{tabular}
\caption{OLS Regression results for LSPs on Ethereum}
\label{OLSETH}
\end{sidewaystable}

\begin{sidewaystable}
    \centering
\begin{tabular}{@{\extracolsep{5pt}}lcccccccc}
\\[-1.8ex]\hline
\hline \\[-1.8ex]
\\[-1.8ex] & stSOL & mSOL & scnSOL & jSOL & stSOL & mSOL & scnSOL & jSOL \\
\hline \\[-1.8ex]
 const & 0.000$^{}$ & 0.000$^{}$ & 0.000$^{}$ & 0.000$^{}$ & 0.004$^{***}$ & -0.001$^{***}$ & -0.001$^{}$ & -0.007$^{***}$ \\
  & (0.000) & (0.000) & (0.001) & (0.000) & (0.000) & (0.001) & (0.001) & (0.000) \\
 $\Delta^{SOL}_{daily}$ & 0.010$^{}$ & 0.025$^{**}$ & 0.072$^{**}$ & 0.024$^{***}$ & 0.016$^{}$ & -0.006$^{}$ & 0.027$^{*}$ & 0.005$^{}$ \\
  & (0.009) & (0.010) & (0.032) & (0.008) & (0.010) & (0.018) & (0.015) & (0.006) \\
 $\sigma_{\text{daily change}}^{SOL}$ & 0.022$^{}$ & 0.010$^{}$ & 0.034$^{}$ & -0.000$^{}$ & -0.018$^{**}$ & -0.010$^{}$ & -0.015$^{}$ & -0.002$^{}$ \\
  & (0.018) & (0.012) & (0.027) & (0.009) & (0.008) & (0.020) & (0.017) & (0.009) \\
 market cap & 0.001$^{}$ & 0.000$^{}$ & 0.001$^{}$ & 0.000$^{}$ & 0.003$^{*}$ & 0.003$^{**}$ & 0.000$^{}$ & 0.001$^{}$ \\
  & (0.001) & (0.001) & (0.002) & (0.001) & (0.002) & (0.001) & (0.003) & (0.001) \\
 $\sigma_{monthly}^{SOL}$ & 0.039$^{}$ & 0.025$^{}$ & 0.032$^{}$ & 0.008$^{}$ & 0.019$^{}$ & 0.024$^{}$ & -0.007$^{}$ & 0.018$^{}$ \\
  & (0.033) & (0.024) & (0.034) & (0.010) & (0.020) & (0.023) & (0.050) & (0.019) \\
 shift1 & & & & & 0.335$^{***}$ & 0.343$^{***}$ & 0.105$^{}$ & 0.346$^{***}$ \\
  & & & & & (0.093) & (0.111) & (0.072) & (0.133) \\
 shift2 & & & & & 0.057$^{}$ & 0.137$^{*}$ & 0.099$^{}$ & 0.156$^{***}$ \\
  & & & & & (0.041) & (0.076) & (0.061) & (0.059) \\
 shift3 & & & & & -0.041$^{}$ & -0.098$^{}$ & 0.103$^{**}$ & 0.090$^{**}$ \\
  & & & & & (0.082) & (0.062) & (0.050) & (0.045) \\
 shift4 & & & & & -0.027$^{}$ & -0.048$^{}$ & 0.043$^{}$ & 0.083$^{*}$ \\
  & & & & & (0.035) & (0.056) & (0.050) & (0.047) \\
 shift5 & & & & & 0.016$^{}$ & 0.021$^{}$ & 0.046$^{}$ & -0.081$^{}$ \\
  & & & & & (0.021) & (0.052) & (0.053) & (0.062) \\
 shift6 & & & & & 0.026$^{}$ & -0.027$^{}$ & 0.027$^{}$ & -0.040$^{}$ \\
  & & & & & (0.034) & (0.079) & (0.046) & (0.047) \\
 total volume & -0.001$^{}$ & 0.000$^{}$ & -0.000$^{}$ & -0.000$^{}$ & -0.003$^{}$ & -0.003$^{**}$ & -0.002$^{}$ & -0.002$^{*}$ \\
  & (0.001) & (0.001) & (0.002) & (0.001) & (0.002) & (0.001) & (0.002) & (0.001) \\
\hline \\[-1.8ex]
 Observations & 550 & 542 & 371 & 501 & 545 & 537 & 366 & 496 \\
 $R^2$ & 0.018 & 0.016 & 0.034 & 0.018 & 0.243 & 0.206 & 0.080 & 0.259 \\
 Adjusted $R^2$ & 0.009 & 0.007 & 0.021 & 0.008 & 0.228 & 0.190 & 0.052 & 0.242 \\

\hline
\hline \\[-1.8ex]
\textit{Note:} & \multicolumn{8}{r}{$^{*}$p$<$0.1; $^{**}$p$<$0.05; $^{***}$p$<$0.01} \\

\end{tabular}
\caption{OLS Regression results for LSPs on Solana}
\label{OLSSOL}
\end{sidewaystable}

\begin{table}[!htbp] \centering
\begin{tabular}{@{\extracolsep{5pt}}lcccccc}
\\[-1.8ex]\hline
\hline \\[-1.8ex]
\\[-1.8ex] & ankrBNB & BNBx & STKBNB & ankrBNB & BNBx & STKBNB \\
\hline \\[-1.8ex]
 const & 0.000$^{}$ & 0.000$^{}$ & 0.000$^{}$ & 0.012$^{***}$ & 0.023$^{***}$ & -0.008$^{***}$ \\
  & (0.000) & (0.000) & (0.000) & (0.001) & (0.000) & (0.000) \\
 $\Delta^{BNB}_{daily}$ & -0.027$^{}$ & -0.000$^{}$ & -0.005$^{}$ & -0.011$^{}$ & -0.008$^{}$ & -0.013$^{}$ \\
  & (0.024) & (0.014) & (0.013) & (0.017) & (0.011) & (0.013) \\
 $\sigma_{\text{daily change}}^{BNB}$ & 0.015$^{}$ & -0.001$^{}$ & 0.003$^{}$ & 0.018$^{}$ & 0.001$^{}$ & 0.000$^{}$ \\
  & (0.021) & (0.005) & (0.006) & (0.013) & (0.004) & (0.004) \\
 market cap & -0.001$^{}$ & 0.006$^{**}$ & 0.002$^{}$ & -0.012$^{}$ & -0.002$^{}$ & -0.003$^{}$ \\
  & (0.004) & (0.003) & (0.005) & (0.009) & (0.004) & (0.004) \\
 $\sigma_{monthly}^{BNB}$ & -0.012$^{}$ & 0.004$^{}$ & 0.005$^{}$ & -0.002$^{}$ & -0.035$^{}$ & 0.003$^{}$ \\
  & (0.044) & (0.017) & (0.029) & (0.093) & (0.027) & (0.045) \\
 shift1 & & & & 0.190$^{***}$ & 0.485$^{***}$ & 0.411$^{***}$ \\
  & & & & (0.064) & (0.057) & (0.054) \\
 shift2 & & & & 0.207$^{**}$ & 0.084$^{}$ & 0.104$^{*}$ \\
  & & & & (0.100) & (0.064) & (0.061) \\
 shift3 & & & & -0.025$^{}$ & 0.104$^{*}$ & 0.153$^{*}$ \\
  & & & & (0.093) & (0.058) & (0.089) \\
 shift4 & & & & -0.049$^{}$ & 0.031$^{}$ & -0.011$^{}$ \\
  & & & & (0.145) & (0.073) & (0.045) \\
 shift5 & & & & 0.076$^{}$ & -0.034$^{}$ & 0.104$^{}$ \\
  & & & & (0.106) & (0.062) & (0.079) \\
 shift6 & & & & 0.015$^{}$ & 0.128$^{*}$ & -0.195$^{***}$ \\
  & & & & (0.130) & (0.069) & (0.066) \\
 total volume & -0.000$^{}$ & -0.001$^{**}$ & -0.000$^{}$ & -0.001$^{*}$ & -0.001$^{*}$ & -0.000$^{}$ \\
  & (0.001) & (0.000) & (0.001) & (0.001) & (0.000) & (0.001) \\
\hline \\[-1.8ex]
 Observations & 116 & 228 & 243 & 111 & 223 & 238 \\
 $R^2$ & 0.014 & 0.020 & 0.002 & 0.182 & 0.564 & 0.337 \\
 Adjusted $R^2$ & -0.030 & -0.002 & -0.019 & 0.092 & 0.541 & 0.305 \\

\hline
\hline \\[-1.8ex]
\textit{Note:} & \multicolumn{6}{r}{$^{*}$p$<$0.1; $^{**}$p$<$0.05; $^{***}$p$<$0.01} \\

\end{tabular}
\caption{OLS Regression for LSPs on Binance}
\label{OLSBNB}
\end{table}

\end{document}